\documentclass[conference]{IEEEtran}
\IEEEoverridecommandlockouts
\usepackage{cite}
\usepackage{amsmath,amssymb,amsfonts}
\usepackage{algorithmic}
\usepackage{graphicx}
\usepackage{textcomp}
\usepackage{xcolor}
\def\BibTeX{{\rm B\kern-.05em{\sc i\kern-.025em b}\kern-.08em
    T\kern-.1667em\lower.7ex\hbox{E}\kern-.125emX}}

\usepackage{algorithm}

\usepackage{gensymb} 
\usepackage{booktabs}
\usepackage{multirow}

\makeatletter
\newcommand{\linebreakand}{%
\end{@IEEEauthorhalign}
\hfill\mbox{}\par
\mbox{}\hfill\begin{@IEEEauthorhalign}
}
\makeatother

\begin{document}

\title{Facilitating Reinforcement Learning for Process Control Using Transfer Learning: Overview and Perspectives
\thanks{This work was supported in part by the National Key R\&D Program of China under Grant 2022YFB3305900, and the National Science and Technology Council of Taiwan under Grant NSTC 112-2221-E-033-053.}
}

\author{\IEEEauthorblockN{1\textsuperscript{st} Runze Lin}
	\IEEEauthorblockA{\textit{College of Control}\\
		\textit{Science and Engineering}\\
		\textit{Zhejiang University}\\
		Hangzhou, China\\
		rzlin@zju.edu.cn}
	\and
	\IEEEauthorblockN{2\textsuperscript{nd} Junghui Chen}
	\IEEEauthorblockA{\textit{Department of}\\
		\textit{Chemical Engineering}\\
		\textit{Chung-Yuan Christian University}\\
		Taoyuan, Taiwan\\
		jason@wavenet.cycu.edu.tw}
	\and
	\IEEEauthorblockN{3\textsuperscript{rd} Lei Xie}
	\IEEEauthorblockA{\textit{College of Control}\\
		\textit{Science and Engineering}\\
		\textit{Zhejiang University}\\
		Hangzhou, China\\
		leix@iipc.zju.edu.cn}
	\and
	\IEEEauthorblockN{4\textsuperscript{th} Hongye Su}
	\IEEEauthorblockA{\textit{College of Control}\\
		\textit{Science and Engineering}\\
		\textit{Zhejiang University}\\
		Hangzhou, China\\
		hysu@iipc.zju.edu.cn}
}

\maketitle

\bstctlcite{IEEEexample:BSTcontrol}

\begin{abstract}
In the context of Industry 4.0 and smart manufacturing, the field of process industry optimization and control is also undergoing a digital transformation. With the rise of Deep Reinforcement Learning (DRL), its application in process control has attracted widespread attention. However, the extremely low sample efficiency and the safety concerns caused by exploration in DRL hinder its practical implementation in industrial settings. Transfer learning offers an effective solution for DRL, enhancing its generalization and adaptability in multi-mode control scenarios. This paper provides insights into the use of DRL for process control from the perspective of transfer learning. We analyze the challenges of applying DRL in the process industry and the necessity of introducing transfer learning. Furthermore, recommendations and prospects are provided for future research directions on how transfer learning can be integrated with DRL to enhance process control. This paper aims to offer a set of promising, user-friendly, easy-to-implement, and scalable approaches to artificial intelligence-facilitated industrial control for scholars and engineers in the process industry.
\end{abstract}


\section{Introduction}\label{Introduction}

\IEEEPARstart{I}{n} the realm of Industry 4.0 and smart manufacturing, intelligent processes are becoming increasingly crucial in the field of process control. The rise of AI for science and engineering has provided forward-looking guidance for empowering and driving modern process industries with a new paradigm of Deep Reinforcement Learning (DRL). In recent years, there has been a growing focus within the process industry on the intersection of DRL and process control \cite{RN1, RN2, RN12}. Corresponding algorithmic improvements and application research have been developed for both continuous and batch processes \cite{RN20, RN21, RN22, RN23}.

Utilizing DRL for process control represents a promising direction towards achieving intelligent manufacturing and enhancing the level of production automation. However, the demands for safety and stability in the process industry are higher compared to fields such as finance, recommendation, and robotics. This heightened demand is a significant reason why the widespread application of DRL faces considerable challenges in industrial settings. It is worth noting that some scholars have introduced the concept of transfer learning into DRL for process control to enhance its safety and practicality \cite{RN3, RN4, RN5}. The integration of transfer learning is beneficial for process control, offering a more rational and practical solution for industrial production \cite{RN91}.

Unfortunately, there has been a lack of systematic analysis and synthesis regarding how transfer learning should be integrated into DRL in the realm of process control domain, especially concerning the research directions that are worth exploring in the field of process systems engineering.

To address this gap, this paper presents insights and prospects on \textit{Facilitating Reinforcement Learning for Process Control Using Transfer Learning}, aiming to provide valuable reference points for researchers in our field.

\section{Reinforcement Learning for Process Control}\label{RL}
Reinforcement Learning (RL) has gained significant attention in both academia and industry, with recent research highlighting its potential. DRL, in particular, garnered widespread interest after AlphaGo's victory over human players in the game of Go \cite{RN10}. DRL focuses on solving sequential decision-making problems in uncertain environments \cite{RN11}, making it particularly useful for process control, where uncertainties often play a crucial role \cite{RN12}. Mathematically, DRL can be framed as an optimal decision-making algorithm capable of managing system randomness. Furthermore, DRL not only enables the storage of offline-trained policies but also supports adaptability and transfer learning \cite{RN4, RN5}.

In the process industry, Proportional-Integral-Derivative (PID) and Model Predictive Control (MPC) are two widely adopted control algorithms \cite{RN22, RN23}. PID control works by adjusting feedback based on the proportion, integral, and derivative of the error signal, which can result in slower and more passive control with potential issues like oscillations or overshoot. In contrast, MPC is a model-based approach that predicts future system behavior and optimizes control inputs in real time \cite{RN120}. However, MPC requires an accurate system model, which may not be feasible for complex or dynamic systems.

In this context, DRL offers a promising alternative by providing a data-driven, model-free approach. It learns through trial and error by interacting with the environment, making it well-suited for complex systems with unknown or time-varying dynamics. Unlike MPC, which relies on a predictive model, DRL is particularly effective for control problems where system behavior is difficult to model or subject to change over time. This makes DRL a strong candidate for addressing challenges in process control. Recent advancements have led to numerous applications and improvements in DRL-based control for process industries \cite{RN3, RN4, RN5, RN6, RN92, RN91, RN38}.

Formally, DRL is a learning paradigm that differs from traditional supervised and unsupervised learning. Reinforcement Learning (RL) aims to train an agent to maximize long-term, discounted rewards through interaction with its environment, denoted as $E$. Typically, RL tasks are modeled as a Markov Decision Process (MDP) with state space $\cal S$ and action space $\cal A$. At each time step, the agent selects an action $\boldsymbol a_t$ given the current state $\boldsymbol s_t$ according to its policy, transitioning to a new state $\boldsymbol s_{t+1}$, while receiving a reward signal $r_t$ from the environment.

RL policies can be either stochastic or deterministic. A stochastic policy is represented as the conditional probability density of actions, $ \pi_\theta({\boldsymbol a_t} | {\boldsymbol s_t}) : {\cal S} \mapsto {\cal P}({\cal A}) $, where $\theta$ parameterizes the policy. In contrast, a deterministic policy is given by $ \pi_\theta({\boldsymbol a_t} | {\boldsymbol s_t}) : {\cal S} \mapsto {\cal A} $, where the action $\boldsymbol a_t$ is chosen as the maximizer of the probability distribution. The immediate reward, $r_t = r({\boldsymbol s_t}, {\boldsymbol a_t})$, is a scalar function of the state-action pair. The total reward over an episode, $R_t$, is the discounted sum of immediate rewards:
\begin{equation}
{R_t} = \sum\limits_{t = 1}^T {{\gamma ^t}} {r_t} = \sum\limits_{t = 1}^T {{\gamma ^t}} r({\boldsymbol s_t},{\boldsymbol a_t}). \label{eq1}
\end{equation}
The performance objective of the RL agent is to maximize the expected discounted return:
\begin{equation}
J\left( {{\pi _\theta }} \right) = \mathop {\max }\limits_{{\pi _\theta }} {\mathbb{E}_{\boldsymbol{s}\sim\rho^\pi,\boldsymbol{a}\sim\pi _\theta}}\left[ {{R_t}\left| {{{\boldsymbol{s}}_t},{{\boldsymbol{a}}_t}} \right.} \right]. \label{eq2}
\end{equation}
The action-value function $Q^\pi({\boldsymbol s_t}, {\boldsymbol a_t})$ is the expected return when taking action $\boldsymbol a_t$ in state $\boldsymbol s_t$ and following policy $\pi$:
\begin{equation}
{Q^\pi }({\boldsymbol s_t},{\boldsymbol a_t}) = {\mathbb{E}_\pi }\left[ {{R_t}\left| {{\boldsymbol s_t},{\boldsymbol a_t}} \right.} \right]. \label{eq3}
\end{equation}
The Bellman equation provides a recursive way to compute the action-value function and solve the optimization problem in Eq. \eqref{eq2}:
\begin{equation}
{Q^\pi }({\boldsymbol s_t},{\boldsymbol a_t}) = {\mathbb{E}_{{r_t},{\boldsymbol s_{t + 1}}\sim E}}\left[ {r_t + \gamma {\mathbb{E}_{{\boldsymbol a_{t + 1}}\sim \pi }}\left[ {{Q^\pi }({\boldsymbol s_{t + 1}},{\boldsymbol a_{t + 1}})} \right]} \right]. \label{eq4}
\end{equation}

\section{Practical Barriers of DRL in Process Control}\label{Practical}
While DRL has demonstrated impressive success in fields like video games and recommendation systems, where data is abundant and safety concerns are minimal, its application in process manufacturing, especially in chemical process control, faces substantial challenges. The complexities of real-world industrial systems present significant obstacles to the widespread adoption of DRL. Two primary issues impede its practical use: low sample efficiency and poor generalization across varying operating conditions. Chemical processes, in particular, are large-scale systems with slow dynamics, making the training, validation, and deployment of DRL agents both time-consuming and resource-intensive.

One of the most pressing challenges is safety concerns and the high cost of trial-and-error learning. DRL typically requires extensive interaction with the system to learn effective control policies, often involving millions of interactions with the environment. In industrial settings, this can translate into days or even weeks of training, during which time safety risks and financial costs become prohibitive. Conducting real-world training in a chemical process environment is often impractical due to the potential hazards and the need for controlled conditions, making it difficult to apply DRL directly to physical systems in industrial production.

A potential solution to this challenge involves the use of simulation environments and accurate system models for DRL training. While these models can help mitigate safety risks and reduce training costs, developing detailed mechanistic models of complex chemical systems is a labor-intensive and highly specialized task. Moreover, these models often require substantial computational resources to train, which further complicates their use in practice. Even with an accurate model, however, the training process remains challenging and computationally expensive, particularly when attempting to replicate the slow, nonlinear dynamics of real-world processes.

Another critical issue is the generalization ability of DRL agents. DRL agents are highly sensitive to the specific conditions under which they are trained, and small changes in the environment—such as variations in system parameters or disturbances—can lead to significant performance degradation. In process control, where operating conditions can vary over time and across different modes, this lack of robustness poses a major limitation. DRL agents trained in one operating condition may fail to adapt effectively to new conditions, which can undermine their performance in time-varying or highly nonlinear environments.

Furthermore, many existing studies of DRL in process control fail to adequately address the feasibility of its application in industrial settings. They often overlook crucial factors such as safety and generalization, which are critical for the large-scale deployment of DRL in chemical process industries. Without addressing these challenges, the practical application of DRL in process control remains limited. Until these barriers are overcome, DRL's potential in chemical process control will remain constrained, and its use in such fields will be slow to expand.

Therefore, the nest portion of this paper will explore the potential application of DRL in process control from a new perspective of transfer learning.

\section{Perspectives on How Transfer Learning Facilitates Reinforcement Learning for Industrial Process Control}\label{Perspectives}
In the following section, we delve into detailed discussions and envision how transfer learning can be combined with reinforcement learning to better facilitate the next generation of smart manufacturing in process industries.

\subsection{World Model-Empowered RL Training Acceleration}
One of the main challenges in applying RL to chemical process control is the extensive time and data required for training. To address this, world models (such as digital twins) can be used to create an internal simulation of the environment. World models leverage surrogate models to represent the underlying dynamics of the chemical process, which are trained on historical data to generate synthetic transitions, allowing the RL agent to perform “dreamed” and “imaginary” rollouts and refine control policies without interacting with the physical system.

For example, as a crucial component of Industry 4.0 smart manufacturing, digital twins can provide multi-dimensional, high-precision, and fine-grained virtual reality integrated simulation models for DRL transfer learning. Constructing a digital twin of industrial processes would greatly aid in the transfer learning, adaptation, and generalization of RL agents between the source and target domains, as switching between one or more different domains requires a scalable multi-mode process description provided by the digital twin.

Our approach \cite{RN4, RN5}, inspired by model-assisted RL and Dream to Control, significantly accelerates learning by minimizing the reliance on real-time data collection. By leveraging world models, the RL agent can adapt more quickly to changes in system dynamics and operational modes, ensuring greater flexibility and robustness during training.

\subsection{Sim2Real Transfer Learning of RL Agents Using Pre-training + Fine-tuning}
Sim2Real transfer learning is crucial for enabling RL agents to adapt to multi-mode and cross-process control in industrial settings. This allows pre-trained RL models to adjust to new operating conditions or processes with minimal retraining, enhancing their generalization across different tasks. It encompasses methods to minimize the performance degradation of DRL policies when transitioning from the simulated world to the real world \cite{RN13}. Typically, DRL agents experience a significant drop in performance when the environment changes. 

Pre-training RL controllers in the source domain(s), coupled with fine-tuning RL-based control policies in the target domain, represents the most common approach in transferring RL agents. This kind of concept aligns well with the design principles of industrial process controllers. In fact, pre-training + fine-tuning has long been applied in process industry optimization control, and its extension to modern RL demonstrates clear potential. Sim2Real transfer of RL agents, initially emphasized in the robotics field, presents a highly feasible solution for our process control society as well.

\begin{figure}
	\centering
	\includegraphics[width=1.0\linewidth]{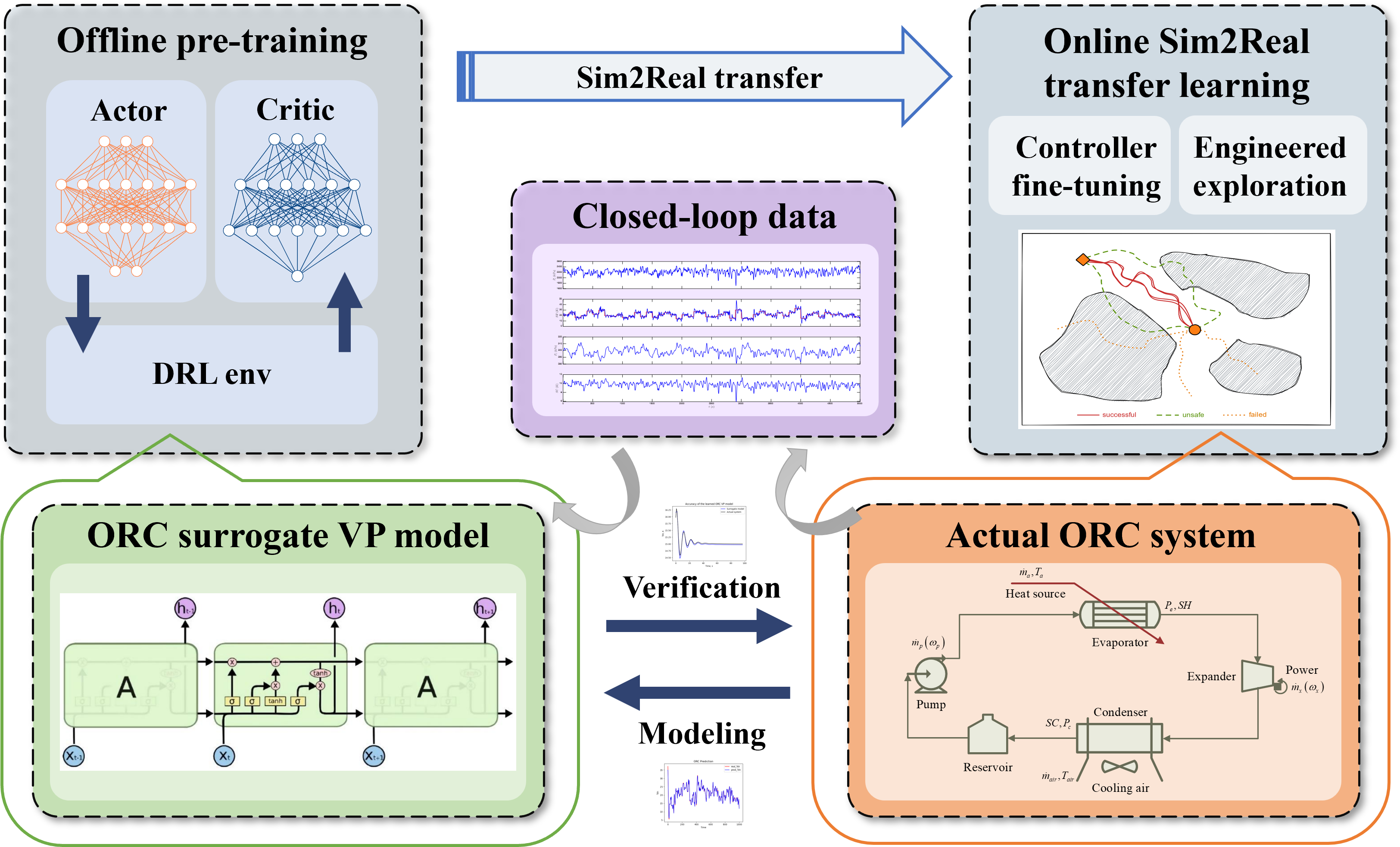}
	\caption{Framework of DRL-based Sim2Real control design \cite{RN3}.}
	\label{fig:fig-1}
\end{figure}

Our previous work \cite{RN3} proposed a simple Sim2Real transfer learning method for the DRL-based controller of a thermodynamic cycle process in the energy field, which addresses the issues of low sample efficiency, time consumption, and safety risks encountered in the interactive training with the actual process control system. The framework is presented in Fig. \ref{fig:fig-1}. We are the first to address the energy system control problem using the concept of DRL Sim2Real transfer learning, innovatively proposing the use of virtual prototypes for offline pre-training and subsequent Sim2Real transfer to the real system.

\subsection{Imitation Learning-Inspired RL Controller Prior Learning}
Learning from demonstrations encompasses various methods for training RL controllers from expert demonstrations, with imitation learning being a common practice. In industrial production, abundant historical closed-loop operation data exists, from which imitation learning can derive controller priors. Subsequently, the initialized RL controllers can undergo transfer learning in actual processes, thereby enhancing the safety of DRL training. For instance, behavior cloning, the simplest imitation learning technique, essentially fits the ``State $s_t$ - Action $a_t$" data pairs from expert trajectories. The results of imitation learning can serve as the foundation for Sim2Real pre-training followed by fine-tuning \cite{RN107}.

\subsection{Apprenticeship Learning for Abstracting Controllers from (Human) Experts}
In the historical operation of process control systems, there exists a vast amount of data generated either by human operators or stored by classical controllers such as PID or MPC. This data is typically underutilized, despite containing valuable information about implicit controller patterns. By extracting potential expert controller behaviors from that kind of historical closed-loop operation data, we can help DRL agents make informed decisions without online exploration.

Apprenticeship Learning \cite{Apprenticeship, RN34}, such as Inverse RL \cite{RN32, RN33, RN19}, can be used to help RL agents learn from human expertise or existing controller(s) without relying on hazardous trial-and-error methods. Inverse RL represents a specialized form of imitation learning, aiming to match the controller to the expert(s) while also recovering the reward function used to rationalize the optimal policy. Compared to general imitation learning, Inverse RL is more suitable for DRL transfer learning as it can infer the closed-loop controller logic hidden within input-output data \cite{RN6}. Estimating the reward function is akin to understanding the intention behind why experts perform such behaviors, rather than simply fitting the controller's behavior. This aids in the expansion and generalization of RL agents from the source domain(s) to the target domain during transfer learning.

In this way, RL agents can be trained offline, learning optimal policies derived from expert human operators or proven control schemes. This method will enable the DRL agent to perform effective “hot/warm start” operations for industrial control systems, reducing the need for costly real-time experimentation and improving both the safety and efficiency of production automation.

This method will create a data-driven controller, leveraging industrial big data and expert knowledge to enhance the intelligence and safety of RL agents for complex processes. Furthermore, the results from offline IRL training could serve as the baseline for subsequent online transfer learning and potentially enable few-shot adaptation.

\subsection{Offline RL towards Safety-Critical Practical Applications}
The primary barrier to the real-world application of DRL in process industries is the significant safety risks associated with online learning and random exploration. Assuming we have accumulated a rich set of data from historical operations in the factory, one straightforward approach to addressing the safety issues in DRL is to leverage these datasets to directly train the process controllers.

As the name suggests, offline RL \cite{RN35, 10078377, he2023surveyofflinemodelbasedreinforcement} involves training the control policy in an entirely offline manner, without the need to interact with the environment, but rather relying on an offline dataset. The advantage of this approach lies in its practicality in transfer learning scenarios, where it serves as the source domain training phase for the RL agents, allowing the controllers to be trained through large datasets that span across state-action spaces.

However, the drawbacks are evident: assuming that the offline dataset encompasses all possible scenarios is unrealistic, as DRL struggles to predict the state-action space it will traverse during its exploration process. Additionally, current research in offline RL mainly focuses on addressing the value function overestimation problem during training. Yet, traditional solutions to this issue tend to introduce conservatism in the resulting control performance \cite{ConservativeQLearning, DoublyMildGeneralization}. Despite these challenges, offline RL remains a promising research field that can DRL towards practical process control applications.

\subsection{Meta-RL or Multi-Task RL for Multi-Mode Control}
Process systems often exhibit multi-mode characteristics, such as changes in setpoints or system parameters, and operating conditions undergoing wide fluctuations. Here, the definition of modes is broad, encompassing any situation that alters the distribution of process systems \cite{RN3, RN4, RN5}. Conventional DRL struggles to adapt well to sudden changes in modes, as it is sensitive to model-plant mismatch, and its generalization performance tends to be poor across different scenarios. A highly promising strategy is meta-RL \cite{2023arXiv230108028B, pmlr-v97-rakelly19a} or multi-task RL \cite{pmlr-v139-sodhani21a, pmlr-v100-yu20a}, capable of training a universal controller to adapt to multiple tasks/modes \cite{RN31, MCCLEMENT2021685}. Given the multi-mode nature of process systems, advanced techniques from data analytics, such as latent space modeling and variational inference, can be integrated to address multi-mode control issues through multi-task learning. Consequently, when the training in the source domains covers some representative modes, DRL agents can quickly adapt to those unseen target domains.

\subsection{Multi-Task Inverse RL for Process Control Using Multi-Mode Historical Big Data}
\begin{figure}
	\centering
	\includegraphics[width=1.0\linewidth]{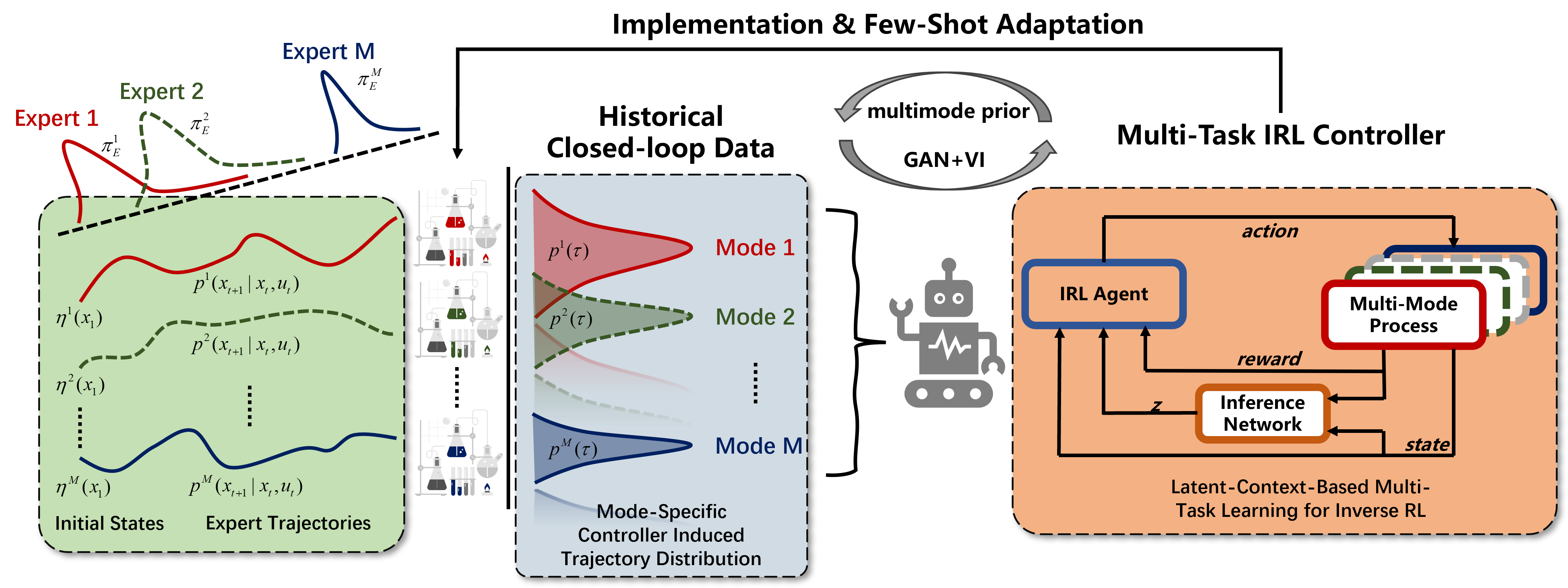}
	\caption{Multi-task inverse RL framework for multi-mode process control design.}
	\label{fig:fig-2}
\end{figure}

This novel idea combines multi-mode controller learning with inverse RL \cite{RN6}, primarily used to recover both reward functions and control policies from large-scale closed-loop data covering various operating modes. The framework is presented in Fig. \ref{fig:fig-2}.

For multi-mode control systems, there exist multiple expert controllers corresponding to different modes, resulting in different distributions of expert trajectories. We can utilize the historical industrial closed-loop big data and multi-task IRL to learn controller prior(s) from mode-specific trajectory distribution through.  Such multi-mode prior(s) will provide a universal controller architecture within the inverse RL framework, facilitating cross-mode few-shot adaptation in the implementation process.

\subsection{Model-based RL or MPC-based RL for Faster Adaptation}
Model-based RL (MBRL) \cite{MBRL} can also serve as a transfer learning solution for RL-based process control. MBRL offers significant advantages for process industry optimization by combining the benefits of system modeling with the adaptability of RL. By incorporating a learned or predefined model of the process, MBRL improves sample efficiency, enabling faster convergence to optimal control policies with fewer real-world interactions. MBRL can also facilitate transfer learning, as the world model captures system dynamics that can be applied across different domains, reducing the need to retrain models for similar processes. This enhances scalability, enabling the deployment of robust, adaptive control strategies across diverse production lines or plants.

Considering the dominant role of MPC in industrial applications, recent research has begun exploring the potential integration of MPC with DRL. Both methods have their own strengths and weaknesses, yet they complement each other effectively. On one hand, MPC can account for uncertainty, system complexity, and long-term prediction horizons, with its performance heavily dependent on the accuracy of the process model used \cite{RN7}. On the other hand, DRL can naturally handle uncertainty in complex systems and manage infinite prediction horizons. However, DRL struggles with constraint satisfaction and lacks interpretability, while MPC provides safety guarantees and transparency. In recent years, some researchers have proposed RL-based MPC, which combines the stability of MPC with the autonomous exploration of RL. Through detailed mathematical derivations, equivalence conditions between nonlinear MPC and RL have been demonstrated \cite{RN37}, and the concept of safe RL using robust MPC has been introduced \cite{RN36}. In this way, the objective function of MPC and the value function of DRL can be integrated \cite{RN87, RN80}, ensuring constraint satisfaction and delivering performance similar to that of MPC, while enabling continuous learning and uncertainty handling.

\subsection{Physics-Informed RL for Incorporating Prior/Physical Knowledge into controllers}
Broadly speaking, any RL algorithm that integrates physical information, such as known laws of physics, system dynamics, or empirical process models, can be categorized as physics-informed RL. By embedding domain-specific knowledge into the learning process, it significantly improves the efficiency and accuracy of the agent’s learning, particularly in complex systems where data-driven approaches alone may not be sufficient.

In a more specific sense, physics-informed RL refers to the combination of RL with Physics-Informed Neural Networks (PINNs) \cite{pmlr-v211-ramesh23a}. PINNs are neural network architectures that directly incorporate physical laws—such as conservation of mass, energy, or momentum—into their training.  This integration ensures that the learned policies respect the governing physical dynamics of the system, improving both the robustness and realism of the control strategies.

A notable advantage of physics-informed RL is its ability to enhance transfer learning. When transferring a learned policy between domains, the embedded physical knowledge acts as a universal feature, providing a stable foundation for the RL agent even across different operational environments. This makes it easier to adapt policies to new but physically similar systems, reducing the need for extensive retraining.

\section{Conclusion}
This paper introduces a set of groundbreaking approaches to chemical process control, addressing key challenges in both traditional methods and current RL techniques. By combining RL with advanced techniques like transfer learning and apprenticeship learning, this framework has the potential to revolutionize how industrial control systems operate. It paves the way for broader adoption of AI in industries, unlocking new levels of efficiency, sustainability, and safety. With the aim of advancing AI-driven technologies in the chemical industry, the ideas proposed by this paper will push the boundaries of what is possible in modern chemical process control, helping to shape the future of industrial automation.


\bibliographystyle{IEEEtran}
\bibliography{ref}

\begin{thebibliography}{10}
\providecommand{\url}[1]{#1}
\csname url@samestyle\endcsname
\providecommand{\newblock}{\relax}
\providecommand{\bibinfo}[2]{#2}
\providecommand{\BIBentrySTDinterwordspacing}{\spaceskip=0pt\relax}
\providecommand{\BIBentryALTinterwordstretchfactor}{4}
\providecommand{\BIBentryALTinterwordspacing}{\spaceskip=\fontdimen2\font plus
\BIBentryALTinterwordstretchfactor\fontdimen3\font minus
  \fontdimen4\font\relax}
\providecommand{\BIBforeignlanguage}[2]{{%
\expandafter\ifx\csname l@#1\endcsname\relax
\typeout{** WARNING: IEEEtran.bst: No hyphenation pattern has been}%
\typeout{** loaded for the language `#1'. Using the pattern for}%
\typeout{** the default language instead.}%
\else
\language=\csname l@#1\endcsname
\fi
#2}}
\providecommand{\BIBdecl}{\relax}
\BIBdecl

\bibitem{RN1}
\BIBentryALTinterwordspacing
J.~Shin, T.~A. Badgwell, K.-H. Liu, and J.~H. Lee, ``Reinforcement learning –
  overview of recent progress and implications for process control,''
  \emph{Computers \& Chemical Engineering}, vol. 127, pp. 282--294, 2019.
  [Online]. Available:
  \url{https://www.sciencedirect.com/science/article/pii/S0098135419300754}
\BIBentrySTDinterwordspacing

\bibitem{RN2}
\BIBentryALTinterwordspacing
R.~Nian, J.~Liu, and B.~Huang, ``A review on reinforcement learning:
  Introduction and applications in industrial process control,''
  \emph{Computers \& Chemical Engineering}, vol. 139, p. 106886, 2020.
  [Online]. Available:
  \url{https://www.sciencedirect.com/science/article/pii/S0098135420300557}
\BIBentrySTDinterwordspacing

\bibitem{RN12}
\BIBentryALTinterwordspacing
H.~Yoo, H.~E. Byun, D.~Han, and J.~H. Lee, ``Reinforcement learning for batch
  process control: Review and perspectives,'' \emph{Annual Reviews in Control},
  vol.~52, pp. 108--119, 2021. [Online]. Available:
  \url{https://www.sciencedirect.com/science/article/pii/S136757882100081X}
\BIBentrySTDinterwordspacing

\bibitem{RN20}
\BIBentryALTinterwordspacing
S.~Spielberg, A.~Tulsyan, N.~P. Lawrence, P.~D. Loewen, and
  R.~Bhushan~Gopaluni, ``Toward self-driving processes: A deep reinforcement
  learning approach to control,'' \emph{AIChE Journal}, vol.~65, no.~10, p.
  e16689, 2019. [Online]. Available:
  \url{https://aiche.onlinelibrary.wiley.com/doi/abs/10.1002/aic.16689}
\BIBentrySTDinterwordspacing

\bibitem{RN21}
\BIBentryALTinterwordspacing
P.~Petsagkourakis, I.~O. Sandoval, E.~Bradford, D.~Zhang, and E.~A. del
  Rio-Chanona, ``Reinforcement learning for batch bioprocess optimization,''
  \emph{Computers \& Chemical Engineering}, vol. 133, p. 106649, 2020.
  [Online]. Available:
  \url{http://www.sciencedirect.com/science/article/pii/S0098135419304168}
\BIBentrySTDinterwordspacing

\bibitem{RN22}
\BIBentryALTinterwordspacing
N.~P. Lawrence, M.~G. Forbes, P.~D. Loewen, D.~G. McClement, J.~U. Backström,
  and R.~B. Gopaluni, ``Deep reinforcement learning with shallow controllers:
  An experimental application to {PID} tuning,'' \emph{Control Engineering
  Practice}, vol. 121, p. 105046, 2022. [Online]. Available:
  \url{https://www.sciencedirect.com/science/article/pii/S0967066121002963}
\BIBentrySTDinterwordspacing

\bibitem{RN23}
\BIBentryALTinterwordspacing
O.~Dogru, K.~Velswamy, F.~Ibrahim, Y.~Wu, A.~S. Sundaramoorthy, B.~Huang,
  S.~Xu, M.~Nixon, and N.~Bell, ``Reinforcement learning approach to autonomous
  {PID} tuning,'' \emph{Computers \& Chemical Engineering}, vol. 161, p.
  107760, 2022. [Online]. Available:
  \url{https://www.sciencedirect.com/science/article/pii/S0098135422001016}
\BIBentrySTDinterwordspacing

\bibitem{RN3}
\BIBentryALTinterwordspacing
R.~Lin, Y.~Luo, X.~Wu, J.~Chen, B.~Huang, H.~Su, and L.~Xie, ``Surrogate
  empowered {Sim2Real} transfer of deep reinforcement learning for {ORC}
  superheat control,'' \emph{Applied Energy}, vol. 356, p. 122310, 2024.
  [Online]. Available:
  \url{https://www.sciencedirect.com/science/article/pii/S0306261923016744}
\BIBentrySTDinterwordspacing

\bibitem{RN4}
\BIBentryALTinterwordspacing
R.~Lin, J.~Chen, L.~Xie, and H.~Su, ``Accelerating reinforcement learning with
  case-based model-assisted experience augmentation for process control,''
  \emph{Neural Networks}, vol. 158, pp. 197--215, 2023. [Online]. Available:
  \url{https://www.sciencedirect.com/science/article/pii/S0893608022004129}
\BIBentrySTDinterwordspacing

\bibitem{RN5}
R.~Lin, J.~Chen, L.~Xie, and H.~Su, ``Accelerating reinforcement learning with
  local data enhancement for process control,'' in \emph{2021 China Automation
  Congress (CAC)}, 2021, Conference Proceedings, pp. 5690--5695.

\bibitem{RN91}
R.~{Lin}, J.~{Chen}, L.~{Xie}, H.~{Su}, and B.~{Huang}, ``{Facilitating
  Reinforcement Learning for Process Control Using Transfer Learning:
  Perspectives},'' \emph{arXiv e-prints}, p. arXiv:2404.00247, Mar. 2024.

\bibitem{RN10}
\BIBentryALTinterwordspacing
V.~Mnih, K.~Kavukcuoglu, D.~Silver, A.~A. Rusu, J.~Veness, M.~G. Bellemare,
  A.~Graves, M.~Riedmiller, A.~K. Fidjeland, G.~Ostrovski, S.~Petersen,
  C.~Beattie, A.~Sadik, I.~Antonoglou, H.~King, D.~Kumaran, D.~Wierstra,
  S.~Legg, and D.~Hassabis, ``Human-level control through deep reinforcement
  learning,'' \emph{Nature}, vol. 518, no. 7540, pp. 529--533, 2015. [Online].
  Available: \url{https://doi.org/10.1038/nature14236}
\BIBentrySTDinterwordspacing

\bibitem{RN11}
\BIBentryALTinterwordspacing
M.~I. Jordan and T.~M. Mitchell, ``Machine learning: Trends, perspectives, and
  prospects,'' \emph{Science}, vol. 349, no. 6245, pp. 255--260, 2015.
  [Online]. Available:
  \url{https://www.science.org/doi/abs/10.1126/science.aaa8415}
\BIBentrySTDinterwordspacing

\bibitem{RN120}
\BIBentryALTinterwordspacing
H.~Hassanpour, X.~Wang, B.~Corbett, and P.~Mhaskar, ``A practically
  implementable reinforcement learning-based process controller design,''
  \emph{AIChE Journal}, vol.~70, no.~1, p. e18245, 2024. [Online]. Available:
  \url{https://aiche.onlinelibrary.wiley.com/doi/abs/10.1002/aic.18245}
\BIBentrySTDinterwordspacing

\bibitem{RN6}
\BIBentryALTinterwordspacing
R.~Lin, J.~Chen, B.~Huang, L.~Xie, and H.~Su, \emph{Developing Purely
  Data-Driven Multi-Mode Process Controllers Using Inverse Reinforcement
  Learning}.\hskip 1em plus 0.5em minus 0.4em\relax Elsevier, 2024, vol.~53,
  pp. 2731--2736. [Online]. Available:
  \url{https://www.sciencedirect.com/science/article/pii/B9780443288241504567}
\BIBentrySTDinterwordspacing

\bibitem{RN92}
L.~Zhang, R.~Lin, L.~Xie, W.~Dai, and H.~Su, ``Event-triggered constrained
  optimal control for organic rankine cycle systems via safe reinforcement
  learning,'' \emph{IEEE Transactions on Neural Networks and Learning Systems},
  vol.~35, no.~5, pp. 7126--7137, 2024.

\bibitem{RN38}
H.~Chang, Q.~Chen, R.~Lin, Y.~Shi, L.~Xie, and H.~Su, ``Controlling pressure of
  gas pipeline network based on mixed proximal policy optimization,'' in
  \emph{2022 China Automation Congress (CAC)}, Conference Proceedings, pp.
  4642--4647.

\bibitem{RN13}
W.~Zhao, J.~P. Queralta, and T.~Westerlund, ``{Sim-to-Real} transfer in deep
  reinforcement learning for robotics: A survey,'' in \emph{2020 IEEE Symposium
  Series on Computational Intelligence (SSCI)}, Conference Proceedings, pp.
  737--744.

\bibitem{RN107}
\BIBentryALTinterwordspacing
Y.~Nakagawa, H.~Ono, Y.~Hazui, and S.~Arai, ``Imitation of piping warm-up
  operation and estimation of operational intention by inverse reinforcement
  learning,'' \emph{Journal of Process Control}, vol. 122, pp. 41--48, 2023.
  [Online]. Available:
  \url{https://www.sciencedirect.com/science/article/pii/S0959152422002335}
\BIBentrySTDinterwordspacing

\bibitem{Apprenticeship}
\BIBentryALTinterwordspacing
P.~Abbeel and A.~Y. Ng, ``Apprenticeship learning via inverse reinforcement
  learning,'' in \emph{Proceedings of the Twenty-First International Conference
  on Machine Learning}, ser. ICML '04.\hskip 1em plus 0.5em minus 0.4em\relax
  New York, NY, USA: Association for Computing Machinery, 2004, p.~1. [Online].
  Available: \url{https://doi.org/10.1145/1015330.1015430}
\BIBentrySTDinterwordspacing

\bibitem{RN34}
\BIBentryALTinterwordspacing
M.~Mowbray, R.~Smith, E.~A. Del Rio-Chanona, and D.~Zhang, ``Using process data
  to generate an optimal control policy via apprenticeship and reinforcement
  learning,'' \emph{AIChE Journal}, vol.~67, no.~9, p. e17306, 2021. [Online].
  Available:
  \url{https://aiche.onlinelibrary.wiley.com/doi/abs/10.1002/aic.17306}
\BIBentrySTDinterwordspacing

\bibitem{RN32}
S.~{Levine}, ``{Reinforcement Learning and Control as Probabilistic Inference:
  Tutorial and Review},'' \emph{arXiv e-prints}, p. arXiv:1805.00909, May 2018.

\bibitem{RN33}
J.~{Fu}, K.~{Luo}, and S.~{Levine}, ``{Learning Robust Rewards with Adversarial
  Inverse Reinforcement Learning},'' \emph{arXiv e-prints}, p.
  arXiv:1710.11248, Oct. 2017.

\bibitem{RN19}
\BIBentryALTinterwordspacing
N.~Ab~Azar, A.~Shahmansoorian, and M.~Davoudi, ``From inverse optimal control
  to inverse reinforcement learning: A historical review,'' \emph{Annual
  Reviews in Control}, vol.~50, pp. 119--138, 2020. [Online]. Available:
  \url{https://www.sciencedirect.com/science/article/pii/S1367578820300511}
\BIBentrySTDinterwordspacing

\bibitem{RN35}
S.~{Levine}, A.~{Kumar}, G.~{Tucker}, and J.~{Fu}, ``{Offline Reinforcement
  Learning: Tutorial, Review, and Perspectives on Open Problems},'' \emph{arXiv
  e-prints}, p. arXiv:2005.01643, May 2020.

\bibitem{10078377}
R.~Figueiredo~Prudencio, M.~R. O.~A. Maximo, and E.~L. Colombini, ``A survey on
  offline reinforcement learning: Taxonomy, review, and open problems,''
  \emph{IEEE Transactions on Neural Networks and Learning Systems}, vol.~35,
  no.~8, pp. 10\,237--10\,257, 2024.

\bibitem{he2023surveyofflinemodelbasedreinforcement}
\BIBentryALTinterwordspacing
H.~He, ``A survey on offline model-based reinforcement learning,'' 2023.
  [Online]. Available: \url{https://arxiv.org/abs/2305.03360}
\BIBentrySTDinterwordspacing

\bibitem{ConservativeQLearning}
A.~Kumar, A.~Zhou, G.~Tucker, and S.~Levine, ``Conservative {Q}-learning for
  offline reinforcement learning,'' in \emph{Proceedings of the 34th
  International Conference on Neural Information Processing Systems}, ser. NIPS
  '20.\hskip 1em plus 0.5em minus 0.4em\relax Red Hook, NY, USA: Curran
  Associates Inc., 2020.

\bibitem{DoublyMildGeneralization}
Y.~Mao, Q.~Wang, Y.~Qu, Y.~Jiang, and X.~Ji, ``Doubly mild generalization for
  offline reinforcement learning,'' in \emph{Advances in Neural Information
  Processing Systems 38: Annual Conference on Neural Information Processing
  Systems 2024, NeurIPS 2024, Vancouver, BC, Canada, December 10 - 15, 2024},
  A.~Globersons, L.~Mackey, D.~Belgrave, A.~Fan, U.~Paquet, J.~M. Tomczak, and
  C.~Zhang, Eds., 2024.

\bibitem{2023arXiv230108028B}
J.~{Beck}, R.~{Vuorio}, E.~{Zheran Liu}, Z.~{Xiong}, L.~{Zintgraf}, C.~{Finn},
  and S.~{Whiteson}, ``{A Survey of Meta-Reinforcement Learning},'' \emph{arXiv
  e-prints}, p. arXiv:2301.08028, Jan. 2023.

\bibitem{pmlr-v97-rakelly19a}
\BIBentryALTinterwordspacing
K.~Rakelly, A.~Zhou, C.~Finn, S.~Levine, and D.~Quillen, ``Efficient off-policy
  meta-reinforcement learning via probabilistic context variables,'' in
  \emph{Proceedings of the 36th International Conference on Machine Learning},
  ser. Proceedings of Machine Learning Research, K.~Chaudhuri and
  R.~Salakhutdinov, Eds., vol.~97.\hskip 1em plus 0.5em minus 0.4em\relax PMLR,
  09--15 Jun 2019, pp. 5331--5340. [Online]. Available:
  \url{https://proceedings.mlr.press/v97/rakelly19a.html}
\BIBentrySTDinterwordspacing

\bibitem{pmlr-v139-sodhani21a}
\BIBentryALTinterwordspacing
S.~Sodhani, A.~Zhang, and J.~Pineau, ``Multi-task reinforcement learning with
  context-based representations,'' in \emph{Proceedings of the 38th
  International Conference on Machine Learning}, ser. Proceedings of Machine
  Learning Research, M.~Meila and T.~Zhang, Eds., vol. 139.\hskip 1em plus
  0.5em minus 0.4em\relax PMLR, 18--24 Jul 2021, pp. 9767--9779. [Online].
  Available: \url{https://proceedings.mlr.press/v139/sodhani21a.html}
\BIBentrySTDinterwordspacing

\bibitem{pmlr-v100-yu20a}
\BIBentryALTinterwordspacing
T.~Yu, D.~Quillen, Z.~He, R.~Julian, K.~Hausman, C.~Finn, and S.~Levine,
  ``{Meta-World}: A benchmark and evaluation for multi-task and meta
  reinforcement learning,'' in \emph{Proceedings of the Conference on Robot
  Learning}, ser. Proceedings of Machine Learning Research, L.~P. Kaelbling,
  D.~Kragic, and K.~Sugiura, Eds., vol. 100.\hskip 1em plus 0.5em minus
  0.4em\relax PMLR, 30 Oct--01 Nov 2020, pp. 1094--1100. [Online]. Available:
  \url{https://proceedings.mlr.press/v100/yu20a.html}
\BIBentrySTDinterwordspacing

\bibitem{RN31}
\BIBentryALTinterwordspacing
D.~G. McClement, N.~P. Lawrence, J.~U. Backström, P.~D. Loewen, M.~G. Forbes,
  and R.~B. Gopaluni, ``Meta-reinforcement learning for the tuning of {PI}
  controllers: An offline approach,'' \emph{Journal of Process Control}, vol.
  118, pp. 139--152, 2022. [Online]. Available:
  \url{https://www.sciencedirect.com/science/article/pii/S0959152422001445}
\BIBentrySTDinterwordspacing

\bibitem{MCCLEMENT2021685}
\BIBentryALTinterwordspacing
D.~G. McClement, N.~P. Lawrence, P.~D. Loewen, M.~G. Forbes, J.~U. Backström,
  and R.~B. Gopaluni, ``A meta-reinforcement learning approach to process
  control,'' \emph{IFAC-PapersOnLine}, vol.~54, no.~3, pp. 685--692, 2021, 16th
  IFAC Symposium on Advanced Control of Chemical Processes ADCHEM 2021.
  [Online]. Available:
  \url{https://www.sciencedirect.com/science/article/pii/S2405896321010958}
\BIBentrySTDinterwordspacing

\bibitem{MBRL}
\BIBentryALTinterwordspacing
T.~M. Moerland, J.~Broekens, A.~Plaat, and C.~M. Jonker, ``Model-based
  reinforcement learning: A survey,'' \emph{Found. Trends Mach. Learn.},
  vol.~16, no.~1, p. 1–118, Jan. 2023. [Online]. Available:
  \url{https://doi.org/10.1561/2200000086}
\BIBentrySTDinterwordspacing

\bibitem{RN7}
\BIBentryALTinterwordspacing
Y.~Shi, R.~Lin, X.~Wu, Z.~Zhang, P.~Sun, L.~Xie, and H.~Su, ``Dual-mode fast
  {DMC} algorithm for the control of {ORC} based waste heat recovery system,''
  \emph{Energy}, vol. 244, p. 122664, 2022. [Online]. Available:
  \url{https://www.sciencedirect.com/science/article/pii/S0360544221029133}
\BIBentrySTDinterwordspacing

\bibitem{RN37}
\BIBentryALTinterwordspacing
S.~Gros and M.~Zanon, ``Data-driven economic {NMPC} using reinforcement
  learning,'' \emph{IEEE Transactions on Automatic Control}, vol.~65, no.~2,
  pp. 636--648, 2020. [Online]. Available:
  \url{https://ieeexplore.ieee.org/document/8701462}
\BIBentrySTDinterwordspacing

\bibitem{RN36}
M.~Zanon and S.~Gros, ``Safe reinforcement learning using robust {MPC},''
  \emph{IEEE Transactions on Automatic Control}, pp. 1--1, 2020.

\bibitem{RN87}
\BIBentryALTinterwordspacing
T.~H. Oh, H.~M. Park, J.~W. Kim, and J.~M. Lee, ``Integration of reinforcement
  learning and model predictive control to optimize semi-batch bioreactor,''
  \emph{AIChE Journal}, vol.~68, no.~6, p. e17658, 2022. [Online]. Available:
  \url{https://aiche.onlinelibrary.wiley.com/doi/abs/10.1002/aic.17658}
\BIBentrySTDinterwordspacing

\bibitem{RN80}
\BIBentryALTinterwordspacing
J.~Arroyo, C.~Manna, F.~Spiessens, and L.~Helsen, ``Reinforced model predictive
  control ({RL-MPC}) for building energy management,'' \emph{Applied Energy},
  vol. 309, p. 118346, 2022. [Online]. Available:
  \url{https://www.sciencedirect.com/science/article/pii/S0306261921015932}
\BIBentrySTDinterwordspacing

\bibitem{pmlr-v211-ramesh23a}
\BIBentryALTinterwordspacing
A.~Ramesh and B.~Ravindran, ``Physics-informed model-based reinforcement
  learning,'' in \emph{Proceedings of The 5th Annual Learning for Dynamics and
  Control Conference}, ser. Proceedings of Machine Learning Research, N.~Matni,
  M.~Morari, and G.~J. Pappas, Eds., vol. 211.\hskip 1em plus 0.5em minus
  0.4em\relax PMLR, 15--16 Jun 2023, pp. 26--37. [Online]. Available:
  \url{https://proceedings.mlr.press/v211/ramesh23a.html}
\BIBentrySTDinterwordspacing

\end{thebibliography}


\end{document}